\documentclass[twocolumn,showpacs,preprintnumbers,amsmath,amssymb]{revtex4}

\usepackage{graphicx}
\usepackage{dcolumn}
\usepackage{bm}

\begin{document}


\title{Role of Frustration and Dimensionality \\
in the Hubbard Model on the Stacked Square Lattice: \\
Variational Cluster Approach}

\author{Toshihiko Yoshikawa}
 \email{yoshikawa@hosi.phys.s.u-tokyo.ac.jp}
\author{Masao Ogata}%
\affiliation{%
Department of Physics, University of Tokyo, Hongo, Bunkyo-ku, Tokyo 113-0033, Japan
}%

\date{\today}

\begin{abstract}
Using variational cluster approach, we study influence of frustration and dimensionality
on magnetic properties in the ground state of Hubbard model on a stacked square lattice in the large $U$ region (U/t=10),
by changing the next-nearest-neighbor hopping, $t^{\prime}$, and the interlayer hopping, $t_{\perp}$.
For small $t_{\perp}<t^{\ast}_{\perp}$ with $t^{\ast}_{\perp}/t{\approx}0.44$,
antiferromagnetic long-range order appears at small $t^{\prime}<t^{\prime}_{c_{1}}$, 
and collinear magnetic long-range order at large $t^{\prime}>t^{\prime}_{c_{2}}$. 
They are separated by a paramagnetic Mott insulating state which appears in the parameter region $t^{\prime}_{c_{1}}<t^{\prime}<t^{\prime}_{c_{2}}$.
For large $t_{\perp}>t^{\ast}_{\perp}$, the paramagnetic Mott insulating state disappears 
and a direct transition between antiferromagnetic state and collinear magnetic state occurs at $t^{\prime}=t^{\prime}_{c_{3}}$.
We also find that the transition from the antiferromagnetic state to the paramagnetic Mott insulating state is of the second-order 
and that from the collinear magnetic state to the paramagnetic Mott insulating or the antiferromagnetic state is of the first-order.
\end{abstract}

\pacs{71.10.Fd, 75.10.Jm, 75.40.Mg}
\maketitle

\section{Introduction} 

Geometrical frustration with strong electronic correlations is one of the main issues in modern condensed-matter physics.
In connection with recent experimental studies of frustrated quantum magnets such as those on triangular, kagom${\acute{\rm e}}$, spinel and pyrochlore lattices
{\cite{ramirez, greedan, shimizu, tamura, hiroi, taguchi, wiebe}}
as well as on triangular structure of $^{3}$He on graphite 
{\cite{ishida}},
Hubbard model on lattices with geometrically frustrated structures have been intensively studied
{\cite{kino, kuroki, morita, parcollet, watanabe, kyung, sahebsara, kyung2, koretsune, imai, bulut, ohashi, mizusaki, yokoyama}}.  
For example, a quantum spin-liquid feature in the organic material ${\kappa}$-(ET)$_{2}$Cu$_{2}$(CN)$_{3}$ stimulated 
the studies on the Hubbard model at half-filling on anisotropic triangular lattices
{\cite{morita, watanabe, kyung, sahebsara, koretsune, mizusaki}}. 
Another example is the two-dimensional Hubbard model with the next-nearest-neighbor hopping, $t^{\prime}$, which will be studied in this paper.
Although geometrical frustration tends to suppress conventional magnetic orderings, the properties of the resultant quantum phases and 
their competitions with other phases are, in many cases, still under discussion.   

\begin{figure}[h]
\begin{tabular}{c}

\begin{minipage}{1.0\linewidth}
\includegraphics[width=\linewidth]{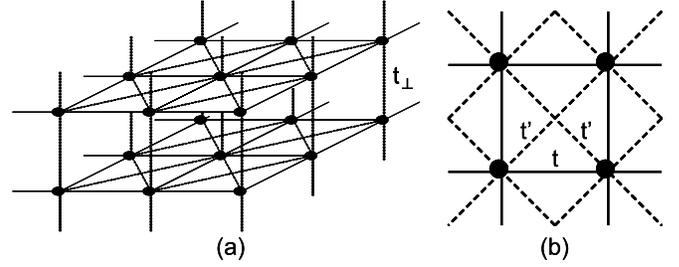}

\caption{The Lattice structure (a) and the hopping integrals $t$, $t^{\prime}$ and $t_{\perp}$ (a)(b) used in this study.
}
\label{fig:lattice-ver010}
\end{minipage}

\end{tabular}
\end{figure}

In this paper, we study electronic properties of the Hubbard model at half-filling on the stacked square lattice as shown in Fig. {\ref{fig:lattice-ver010}}.
The Hamiltonian is given by 
\begin{eqnarray} 
H&=&-{\sum_{l}}({\sum_{{\langle}i,j{\rangle},{\sigma}}}tc^{\dagger}_{i,l,{\sigma}}c_{j,l,{\sigma}}
-{\sum_{[i,j],{\sigma}}}t^{\prime}c^{\dagger}_{i,l,{\sigma}}c_{j,l,{\sigma}}) \nonumber \\
&&-{\sum_{i,l,{\sigma}}}t_{\perp}(c^{\dagger}_{i,l,{\sigma}}c_{i,l+1,{\sigma}}+c^{\dagger}_{i,l,{\sigma}}c_{i,l-1,{\sigma}}) \nonumber \\
&&-{\mu}{\sum_{i,l,{\sigma}}}n_{i,l,{\sigma}}+U{\sum_{i,l}}n_{i,l,{\uparrow}}n_{i,l,{\downarrow}}.
\label{eq:hm1}
\end{eqnarray}
Here $l$ labels the layers.
We consider three different hoppings $t$, $t^{\prime}$ and $t_{\perp}$ where $t$ ($t^{\prime}$) is nearest- (next-nearest-) neighbor hopping in the same layer, 
and $t_{\perp}$ is interlayer hopping. 
Note that signs of $t$, $t^{\prime}$ and $t_{\perp}$ do not affect the results due to particle-hole symmetry for half-filled case.
The operator $c_{i,l,{\sigma}}$ ($c^{\dagger}_{i,l,{\sigma}}$) annihilates (creates) an electron with spin ${\sigma}$ at the site $i$ in the layer $l$,
and $n_{i,l,{\sigma}}$ is the corresponding occupation number operator.
${\mu}$ is the chemical potential, and $U$ is the local Coulomb repulsion.
At half-filling,
the supression of conventional magnetic orders and their competitions are still under discussion. 

In the large $U$ region, the half-filled Hubbard model in Eq. ({\ref{eq:hm1}})
with $t_{\perp}=0$ can be transformed into the two-dimensional $J_{1}$-$J_{2}$ Heisenberg model on the square lattice.
This $J_{1}$-$J_{2}$ model has been studied extensively
{\cite{chandra, dagotto, schulz, richter, siurakshina, bishop, singh, capriotti, sushkov, capriotti2, singh2, roscilde, schmalfus}}.
When $J_{2}=0$ or $t^{\prime}/t=0$, 
it is well accepted that the ground state exhibits antiferromagnetic long-range order 
with the magnetic wave vector ${\bf Q_{\rm 0}}=({\pi},{\pi})$ due to a perfect nesting of fermi surface. 
As $J_{2}$ increases, the antiferromagnetism is increasingly frustrated.
In the limit of $J_{2}/J_{1}={\infty}$,
a collinear magnetic state with the magnetic wave vector ${\bf Q_{\rm 1}}=({\pi},0)$ or $(0,{\pi})$ is realized.
This state consists of two antiferromagnetic long-range orders formed on the two sublattices.
In the small $U$ region,  
Mizusaki $et$ $al.$ recently studied the two-dimensional Hubbard model, Eq. ({\ref{eq:hm1}}) with $t_{\perp}=0$,
using path-integral renormalization group method {\cite{mizusaki}}.
They found a long-period antiferromagnetic-insulator phase with $2{\times}4$ structure for intermediate $t'/t{\approx}0.7$ at large $U/t>7$.
Furthermore, a quantum spin liquid phase with gapless spin excitations 
is found near the Mott transition to paramagnetic metals.
They also claim that there is degeneracy of the ground states
with various total momenta within the whole Brillouin zone
in the quantum spin liquid phase.
Therefore, various unconventional states are expected in the frustrated Hubbard model.

The main problem we would like to study in this paper is the influence of frustration and dimensionality, 
namely that of $t^{\prime}/t$ and $t_{\perp}/t$,
on the magnetic property of the frustrated Hubbard model in Eq. ({\ref{eq:hm1}}). 
First, we will show that a paramagnetic insulating state, similar to that obtained by Mizusaki $et$ $al.$, is obtained 
in a different numerical method for $0.71<t'/t<0.85$ and $U/t=10$.
Then we will study its stability by increasing the three dimensionality ($t_{\perp}$).
Actually, the dimensionality also has strong influence on the property of magnetism
{\cite{schollwock}}.
Although the tendency to order is more pronounced in three-dimensional systems than in low-dimensional ones,
a magnetically disordered phase can be seen in frustrated systems such as a pyrochlore
and a stacked kagom${\acute{\rm e}}$ lattice
{\cite{canals, schmalfus2}}. 
Therefore, high-dimensionalization (adding $t_{\perp}/t$), 
which naively seems to have the competing effect with the effect of $t^{\prime}/t$ on the ordering phenomenon, 
will affect the ground-state property.

The exact diagonalization used in the two-dimensional Hubbard model will not be appropriate 
for the three-dimensional problem under consideration.
Therefore, we use a quantum cluster approach
{\cite{maier}}
to describe strong electron correlations and geometrical frustration.
In recent years there has been a substantial progress in numerical techniques using quantum cluster theories, 
such as cluster extensions  of dynamical mean-field theory (DMFT)
{\cite{georges}},
i.e., 
dynamical cluster approximation (DCA)
{\cite{hettler}}
and cellular DMFT (CDMFT)
{\cite{kotliar}},
or variational cluster approach (VCA)
{\cite{potthoff, dahnken}}.
In this paper, we apply VCA which was proposed recently.
It is based on a self-energy-functional theory (SFT)
{\cite{potthoff-sft}} 
which provides a general variational scheme using dynamical information
from an exactly solvable ``reference system" (in the present case, an isolated cluster) in order to study the infinite-size lattice fermion problem.
It has been recently applied to the broken symmetry phases in the Hubbard model
{\cite{senechal, aichhorn-hightc1, aichhorn-hightc2, aichhorn-hightc3}}
and 
the dimer Hubbard model as a model for the organic materials ${\kappa}$-(ET)$_{2}$X
{\cite{sahebsara}}.

This paper is organized as follows: We start with a brief review of SFT in general (Sec {\ref{sec:sft}}),
and then describe VCA applied in the Hamiltonian ({\ref{eq:hm1}}) (Sec {\ref{sec:vca}}).
Some technical details are addressed in Appendix.
In Sec. {\ref{sec:results}}, our results (energies, order parameters and a phase diagram) are presented and discussed.
Finally, Sec. {\ref{sec:summary}} contains our main conclusions and a summary.

\section{Theoretical background}

In order to describe the VCA used in this paper, it is useful to explain the SFT first.

\subsection{Self-energy-functional theory{\rm {\cite{potthoff-sft}}}}
\label{sec:sft}

For a system with Hamiltonian $H=H_{0}({\bf t})+H_{1}({\bf U})$, where ${\bf t}$ represents the one-particle and ${\bf U}$ the interaction parameters,
the grand potential of the system at temperature, $T$, and chemical potential, ${\mu}$, can be written as a functional of the self-energy ${\bf {\Sigma}}$:
\begin{equation}
{\Omega}_{{\bf t},{\bf U}}[{\bf {\Sigma}}]={\rm Tr}{\verb| |}{\ln}({\bf G}^{-1}_{0,{\bf t}}-{\bf {\Sigma}})^{-1}+F_{\bf U}[{\bf {\Sigma}}],
\label{eq:omega1}
\end{equation}
with the stational property ${\delta}{\Omega}_{{\bf t},{\bf U}}[{\bf {\Sigma}}_{\rm phys}]=0$ for the physical self-energy.
Here, ${\bf G}_{0,{\bf t}}=({\omega}+{\mu}-{\bf t})^{-1}$ is the free Green's function and 
Tr is defined as ${\rm Tr}{\verb| |}{\equiv}{\verb| |}T{\sum_{{\omega}_{n}}}e^{i{\omega}_{n}0^{+}}{\rm tr}$,
with tr being the usual trace and ${\omega}_{n}$ being the Matsubara frequencies, ${\omega}_{n}=(2n+1){\pi}T$, for integer $n$.
$F_{\bf U}[{\bf {\Sigma}}]$ is the Legendre transform of the universal Luttinger-Ward Functional ${\Phi}_{\bf U}[{\bf G}]$
{\cite{luttinger}}.
Because the Luttinger-Ward Functional is defined via an infinite sum of renormalized skeleton diagrams,
the functional dependence ${\Phi}_{\bf U}[{\bf G}]$ or $F_{\bf U}[{\bf {\Sigma}}]$ is not known explicitly.
However, it is important that the functional form $F_{\bf U}[{\bf {\Sigma}}]$ is independent of ${\bf t}$.

Due to this universality of $F_{\bf U}[{\bf {\Sigma}}]$, we have
\begin{equation}
{\Omega}_{{\bf t^{\prime}},{\bf U}}[{\bf {\Sigma}}]={\rm Tr}{\verb| |}{\ln}({\bf G}^{-1}_{0,{\bf t^{\prime}}}-{\bf {\Sigma}})^{-1}+F_{\bf U}[{\bf {\Sigma}}],
\label{eq:omega2}
\end{equation}
for the self-energy functional of a reference system, which is given by a Hamiltonian with the same interaction part ${\bf U}$ but 
modified one-particle parameters ${\bf t^{\prime}}$, i.e. $H^{\prime}=H_{0}({\bf t^{\prime}})+H_{1}({\bf U})$.
By a proper choice of $H_{0}({\bf t^{\prime}})$, the problem posed by the reference system $H^{\prime}$ can be much simpler than the original probrem posed by $H$,
such that the self-energy of the reference system ${\bf {\Sigma}}_{{\bf t^{\prime}},{\bf U}}$ can be computed exactly 
within a certain subspace of parameters ${\bf t^{\prime}}$ which we call a subspace, $S$.
Combining Eqs. ({\ref{eq:omega1}}) and ({\ref{eq:omega2}}), we can eliminate the functional $F_{\bf U}[{\bf {\Sigma}}]$.
Inserting the self-energy of the reference system as a trial self-energy, we obtain 
\begin{eqnarray}
{\Omega}_{{\bf t},{\bf U}}[{\bf {\Sigma}}_{{\bf t^{\prime}},{\bf U}}]&=&
{\Omega}_{{\bf t^{\prime}},{\bf U}}
+{\rm Tr}{\verb| |}{\ln}({\bf G}^{-1}_{0,{\bf t}}-{\bf {\Sigma}}_{{\bf t^{\prime}},{\bf U}})^{-1} \nonumber \\
&&-{\rm Tr}{\verb| |}{\ln}{\bf G}_{{\bf t^{\prime}},{\bf U}},
\label{eq:omega3}
\end{eqnarray}
where ${\Omega}_{{\bf t^{\prime}},{\bf U}}$ and 
${\bf G}_{{\bf t^{\prime}},{\bf U}}=({\bf G}^{-1}_{0,{\bf t^{\prime}}}-{\bf {\Sigma}}_{{\bf t^{\prime}},{\bf U}})^{-1}$ 
are the grand potential and the Green's function of the reference system.
Stationaly points are obtained in the restricted subspace {\it S} of trial self-energy
(${\bf {\Sigma}}_{{\bf t^{\prime}},{\bf U}}{\verb| |}{\in}{\verb| |}{\it S}$).
Varying the trial self-energy in {\it S} is carried out by varying the one-particle parameters ${\bf t^{\prime}}$ of the reference system. 
In addition to the variational parameters, ${\bf t^{\prime}}$, suitably chosen fictitious symmetry-breaking Weiss fields 
can be introduced as variational parameters. By this method, 
normal and off-diagonal long-range order can be studied.
Note that CPT, VCA and CDMFT, which treat decoupled clusters differently, are understood from a unified framework of SFT.

\subsection{Variational cluster approach} 
\label{sec:vca}

VCA is conceptually clear and simple, and much easier to implement numerically.
Therefore, we apply VCA to the Hamiltonian Eq. (\ref{eq:hm1}).
We rewrite Eq. (\ref{eq:hm1}) as 
\begin{equation}
H=H_{0}({\bf t})+H_{1},
\label{eq:hm2}
\end{equation}
where $H_{1}$ is the interaction part,
\begin{equation} 
H_{1}=U{\sum_{i,l}}n_{i,l,{\uparrow}}n_{i,l,{\downarrow}},
\end{equation}
and $H_{0}({\bf t})$ is the rest of the Hamiltonian Eq. (\ref{eq:hm1}).

In the following, we fix the parameter, $U$, as $U/t=10$, and the average particle number as 
${\langle}n_{i,l,{\uparrow}}{\rangle}+{\langle}n_{i,l,{\downarrow}}{\rangle}=1$,
and consider at zero temperature.
The energy scale is set by choosing $t=1$.

\begin{figure}[t]
\begin{tabular}{c}

\begin{minipage}{1.0\linewidth}
\includegraphics[width=\linewidth]{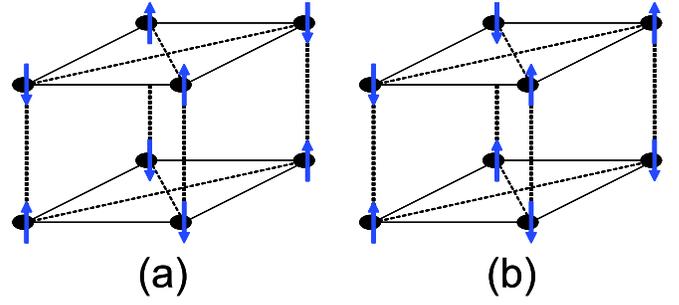}

\caption{Superclusters which tile the three-dimensional lattice (a) with antiferromagnetic long-range order 
and (b) with collinear magnetic long-range order.
Each supercluster consists of two $2{\times}2$ clusters which stack in the $z$-direction.
A major spin at each site is shown by an blue arrow.}
\label{fig:supercluster-ver010}
\end{minipage}

\end{tabular}
\end{figure}

We assume that the Hamiltonian of the reference system $H^{\prime}$ is given by a set of decoupled clusters of a finite size.
For an individual cluster, the Hamiltonian reads
\begin{equation}
H^{\prime}_{\rm cluster}=H^{\prime}_{\rm Hub}+H^{\prime}_{\rm AF}+H^{\prime}_{\rm CM}+H^{\prime}_{\rm local}.
\end{equation} 
Here, $H^{\prime}_{\rm Hub}$ is the Hubbard Hamiltonian in the cluster,
$H^{\prime}_{\rm AF}$ and $H^{\prime}_{\rm CM}$ are two symmetry-breaking terms (Weiss fields),
\begin{equation}
H^{\prime}_{\rm AF}=h^{\prime}_{\rm AF}{\sum_{i,l}}(n_{i,l,{\uparrow}}-n_{i,l,{\downarrow}})e^{i{\bf Q}_{\rm AF}{\cdot}{\bf R}_{i,l}}
\end{equation}
and 
\begin{equation}
H^{\prime}_{\rm CM}=h^{\prime}_{\rm CM}{\sum_{i,l}}(n_{i,l,{\uparrow}}-n_{i,l,{\downarrow}})e^{i{\bf Q}_{\rm CM}{\cdot}{\bf R}_{i,l}},
\end{equation}
where $h^{\prime}_{\rm AF}$ and $h^{\prime}_{\rm CM}$ are the strength of the fields used as variational parameters.
${\bf Q}_{\rm AF}=({\pi},{\pi},{\pi})$ is the antiferromagnetic (AF) wave vector, 
and ${\bf Q}_{\rm CM}=({\pi},0,{\pi})$ is the collinear magnetic (CM) wave vector.
Furthermore, in order to ensure thermodynamic consistency with respect to the average particle number,
the site-independent energy, ${\varepsilon}^{\prime}$, is also treated as a variational parameter
{\cite{aichhorn-hightc1}} which appears in the local term
\begin{equation}
H^{\prime}_{\rm local}={\varepsilon}^{\prime}{\sum_{i,l,{\sigma}}}n_{i,l,{\sigma}}.
\end{equation}
We use a three-dimensional cluster (called as supercluster) with $2{\times}2{\times}2$ sites 
which consists of two $2{\times}2$ clusters stacked in the $z$-direction as shown in Fig. {\ref{fig:supercluster-ver010}}.
and diagonalize the Hamiltonian, $H^{\prime}_{\rm cluster}$.

The trace in Eq. ({\ref{eq:omega3}}) is evaluated accurately as follows.
By converting the frequency integrals to a sum over the poles of the Green's function 
{\cite{potthoff-sft}},
we have
\begin{equation}
{\rm Tr}{\verb| |}{\ln}({\bf G}^{-1}_{0,{\bf t}}-{\bf {\Sigma}}_{{\bf t^{\prime}},{\bf U}})^{-1}
\stackrel{T=0}{=}{\sum_{m}}{\omega}_{m}{\Theta}(-{\omega}_{m})-R
{\label{eq:r1}}
\end{equation} 
and
\begin{equation}
{\rm Tr}{\verb| |}{\ln}{\bf G}_{{\bf t^{\prime}},{\bf U}}
\stackrel{T=0}{=}{\sum_{m}}{\omega}^{\prime}_{m}{\Theta}(-{\omega}^{\prime}_{m})-R.
{\label{eq:r2}}
\end{equation} 
Here ${\Theta}({\omega})$ is the Heaviside step function and
${\omega}^{\prime}_{m}$ are the poles of ${\bf G}_{{\bf t^{\prime}},{\bf U}}$, i.e., 
the one-particle excitation energies of the cluster, ${\omega}^{\prime}_{m}=E_{r}-E_{s}$, which is obtained by exact diagonalization.
Here, we introduce the notation, $m=(r,s)$, to indicate an excitaion between two states $s$ and $r$.
Similarly, ${\omega}_{m}$ are the poles of the Green's function 
$({\bf G}^{-1}_{0,{\bf t}}-{\bf {\Sigma}}_{{\bf t^{\prime}},{\bf U}})^{-1}$.
In Eqs. (\ref{eq:r1}) and (\ref{eq:r2}), 
$R$ represents a contribution due to the poles of the self-energy which cancels out in Eq. ({\ref{eq:omega3}}) and can thus be ignored.
Numerical techniques to calculate ${\omega}_{m}$ and the Green's function of the supercluster are addressed in Appendixes\ref{sec:appendixa} and\ref{sec:appendixb}.

\section{Results}
\label{sec:results}

\subsection{Total energy}

\begin{figure}[h]
\begin{tabular}{c}

\begin{minipage}{1.0\linewidth}
\includegraphics[width=\linewidth]{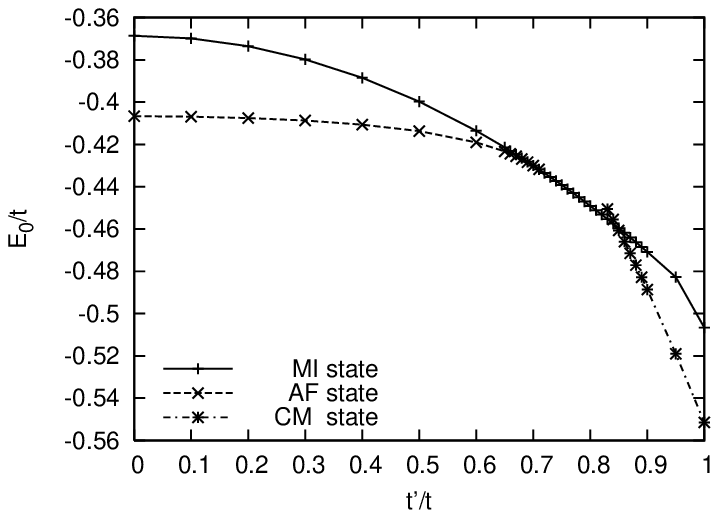}

\caption{$t^{\prime}$ dependence of the total energy per site, $E_{0}$, for $t_{\perp}=0$ obtained in three different states:
paramagnetic Mott insulating (MI) (solid line), antiferromagnetic (AF) (dashed line) and collinear magnetic (CM) states (dash-dotted line).}
\label{fig:energy1}
\end{minipage}
\vspace{2em}
\\
\begin{minipage}{1.0\linewidth}
\includegraphics[width=\linewidth]{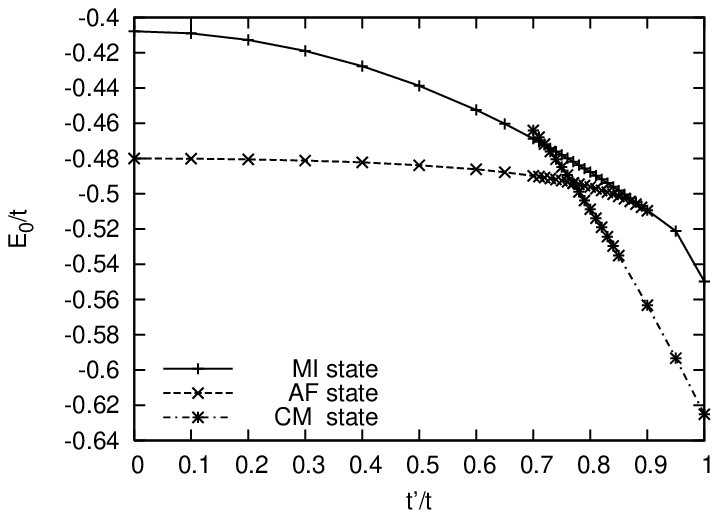}

\caption{Same as in Fig. {\ref{fig:energy1}}, for $t_{\perp}=0.7$.}
\label{fig:energy2}
\end{minipage}
\end{tabular}
\end{figure}

First, let us consider the behavior of the total energy per site $E_{0}$.
Figure {\ref{fig:energy1}} shows the obtained values of $E_{0}$ for $t_{\perp}=0$ as a function of $t^{\prime}$.
In this case, layers are indepenedent of each other.
$E_{0}$'s for paramagnetic Mott insulating (MI) state without magnetic long-range order,
AF and CM states are denoted by solid, dashed and dash-dotted lines, respectively.
Note that these three states are insulating because of gap opening in the total density of states (not shown here).  
Figure {\ref{fig:energy1}} indicates that three different phases appear as a ground state for different values of $t^{\prime}$,
namely that two magnetically long-range ordered states appear at small and large $t^{\prime}$
They are separated by MI state in the parameter region $t^{\prime}_{c_{1}}<t^{\prime}<t^{\prime}_{c_{2}}$,
where $t^{\prime}_{c_{1}}{\approx}0.71$ and $t^{\prime}_{c_{2}}{\approx}0.85$.
$E_{0}$ for MI (solid line) and $E_{0}$ for AF (dashed line)
smoothly coinside as increasing $t^{\prime}$, indicating the second-order phase transition.
On the other hand, 
$E_{0}$ for MI (solid line) and $E_{0}$ for CM (dash-dotted line)
cross each other.
Furthermore, hysteresis behavior is observed at the transition point $t_{c_{2}}$, indicating that
the transition from MI to CM is of the first order.

Figure {\ref{fig:energy2}} shows the same type of data for a rather strong interlayer coupling, $t_{\perp}=0.7$
which indicates that only two different phases appear as a ground state,
namely AF state for $t^{\prime}<t^{\prime}_{c_{3}}$ and CM state for $t^{\prime}>t^{\prime}_{c_{3}}$, respectively, with $t^{\prime}_{c_{3}}{\approx}0.77$.
In this case MI state does not appear and 
the transition from AF to CM is of the first order.
The comparison of Figs. {\ref{fig:energy1}} and {\ref{fig:energy2}} shows that the MI state,
which is stable in a rather wide $t^{\prime}$ region in the strictly two-dimensional system, becomes unstable
by including the three-dimensionality.

\subsection{Order parameters and phase diagram}

\begin{figure}[h]
\begin{tabular}{c}

\begin{minipage}{1.0\linewidth}
\includegraphics[width=\linewidth]{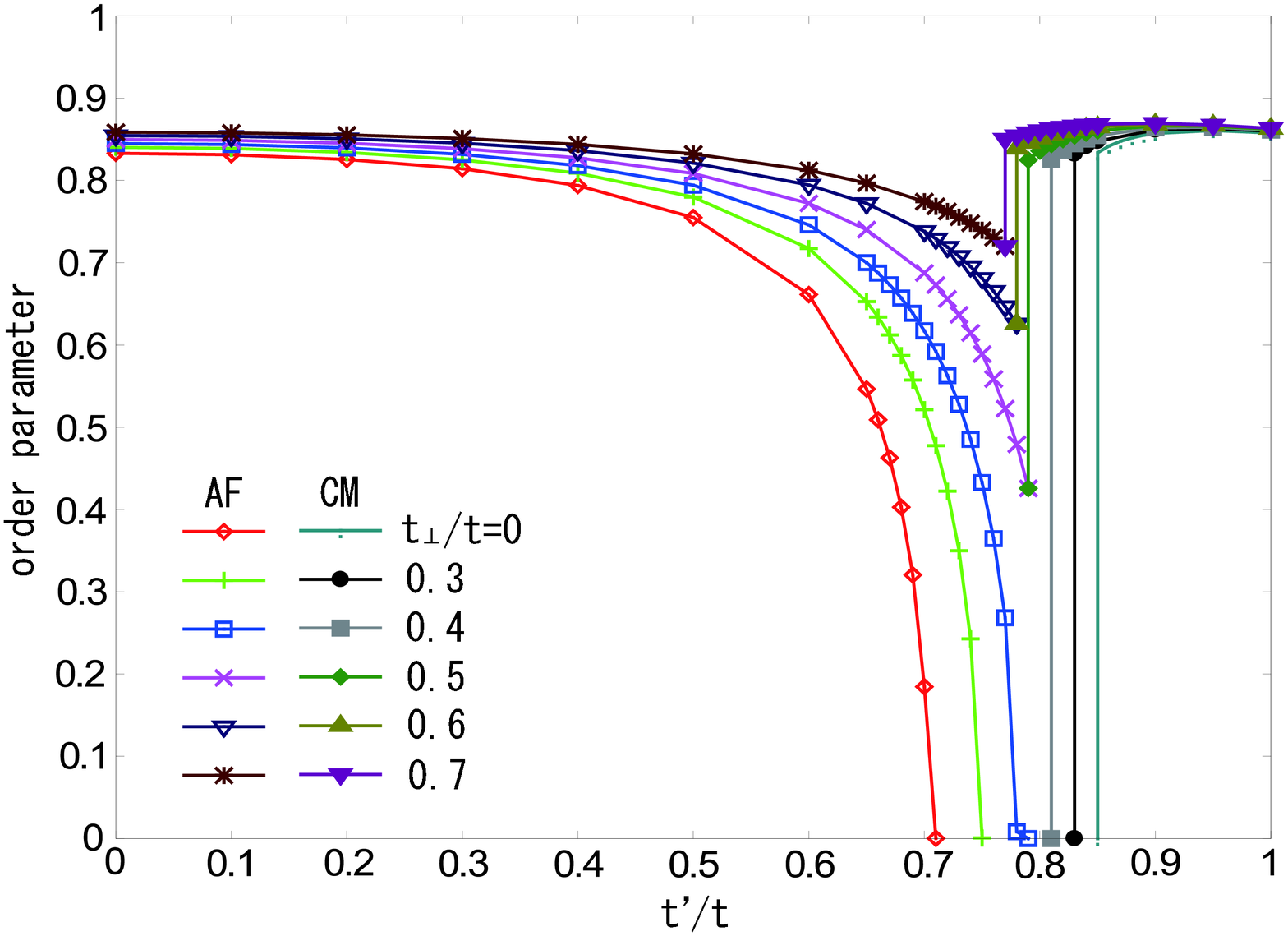}

\caption{$t^{\prime}$ dependence of antiferromagnetic (AF) and collinear magnetic (CM) order parameters for various values of $t_{\perp}$.
At small $t^{\prime}$, AF order parameters for $t_{\perp}/t=0,0.3,0.4,0.5,0.6,0.7$ are shown from bottom to top.
At large $t^{\prime}$, CM order parameters for the same values of $t_{\perp}$ are shown from bottom to top.}
\label{fig:orderparameter-ver020}
\end{minipage}

\end{tabular}
\end{figure}

\begin{figure}[h]
\begin{tabular}{c}

\begin{minipage}{1.0\linewidth}
\includegraphics[width=\linewidth]{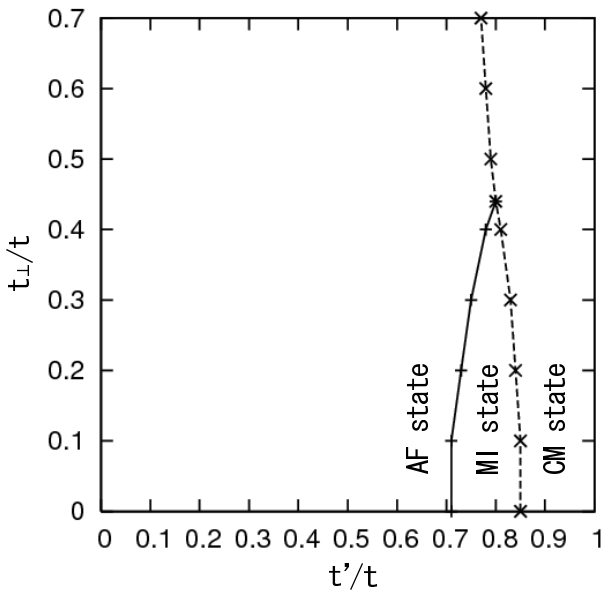}

\caption{Phase diagram in the $t^{\prime}$-$t_{\perp}$ plane.
Transition from antiferromagnetic (AF) state to paramagnetic Mott insulating (MI) state is of the second-order (denoted by solid line).
Transition from collinear magnetic (CM) state to MI or AF state is of the first-order (denoted by dashed line).
}
\label{fig:pd3}
\end{minipage}

\end{tabular}
\end{figure}

Figure {\ref{fig:orderparameter-ver020}} shows the $t^{\prime}$ dependences of 
the AF order parameter at small $t^{\prime}$ and CM order parameter at large $t^{\prime}$. 
In the region where no order parameter appears, MI state is stable as a ground state.
As expected, the order parameters are monotonously increasing with $t_{\perp}$ and
the phase transition points $t^{\prime}_{c_{1}}$, $t^{\prime}_{c_{2}}$ changes as a function of $t_{\perp}$, which is shown in Fig. {\ref{fig:pd3}}.
At $t^{\ast}_{\perp}{\approx}0.44$, the transition points $t^{\prime}_{c_{1}}$ and $t^{\prime}_{c_{2}}$ coinside. 
For $t_{\perp}<t^{\ast}_{\perp}$, AF order parameter vanishes continuously as is typical for the second-order transition.
On the other hand, CM order parameter abruptly drops to zero, which is consistent with the first-order transition.
For $t_{\perp}>t^{\ast}_{\perp}$, the values of AF order parameter and CM order parameter are discontinuously connected, 
indicating the first-order transition.
The obtained phase diagram is summarised in Fig. {\ref{fig:pd3}}, 
where solid line denotes the second-order transition and dashed line denotes the first-order transition.

Here let us compare our results with $t_{\perp}=0$ with those in
the $J_{1}$-$J_{2}$ Heisenberg model on the square lattice.
It has been shown that for $J_{2}/J_{1}<0.4$, the AF state appears and 
for $J_{2}/J_{1}>0.6$, the CM state appears
{\cite{singh2}}.
Since $J_{1}{\approx}\frac{4t^{2}}{U}$ and $J_{2}{\approx}\frac{4t^{{\prime}2}}{U}$,
these critical values correspond to $t^{\prime}/t{\approx}0.63$ and $0.77$
which roughly agree with $t^{\prime}_{c1}$ and $t^{\prime}_{c2}$ obtained in the present method.
For the intermediate region of $0.4<J_{2}/J_{1}<0.6$, no definite conclusion has been drawn on the nature of the ground state of $J_{1}$-$J_{2}$ Heisenberg model.
Possibilities of columnar-dimerized state
{\cite{sushkov, read, sirker}}, 
plaquette singlet state
{\cite{zhitomirsky}}, 
and resonating-valence-bond state have been discussed.
The intermediate state obtained in our method is a nonmagnetic insulator.
The identification of this state is a remaining future problem.

Schmalfu{\ss} {\it et.al.} studied the stacked 
$J_{1}$-$J_{2}$ Heisenberg model using a coupled-cluster method and rotation-invariant Green's function method {\cite{schmalfus}}.
They showed similar phase diagrams as ours, indicating the desappearance of an intermediate magnetically disordered quantum paramagnetic phase
for quite small $J_{\perp}/J_{1}{\approx}0.2-0.3$.
This critical value corresponds to $t_{\perp}/t{\approx}0.45-0.55$ which agree again with Fig. {\ref{fig:pd3}}.

\section{Summary and conclusions} 
\label{sec:summary}

To summarize, we have studied the role of frustration and dimensionality 
in the $t$-$t^{\prime}$ Hubbard model on the stacked square lattice given by the Hamiltonian, Eq. (\ref{eq:hm1}),
at half-filling and at zero temparature, by changing $t^{\prime}$ and $t_{\perp}$.
We have employed recently proposed variational cluster approach (VCA).
Important advantages of VCA are that local and off-site short-range correlations are captured exactly, 
and that broken symmetries are treated with a rigorous dynamical variational principle.   
Therefore, VCA is a powerful method for analyzing magnetic properties of Hubbard model
with strong electron correlations and geometrical frustration. 

Our results show that $t^{\prime}$ destroys AF long-range ordering continuously at small $t^{\prime}$ 
and induces CM long-range ordering at large $t^{\prime}$ with the first-order transition.
There is a MI state at intermediate $t^{\prime}$.   
We also find that $t_{\perp}$ has a tendency to stabilize the magnetic long-range orders and finally MI state vanishes at a critical value of $t_{\perp}$
($t^{\ast}_{\perp}{\approx}0.44$).
Therefore, interesting phenomena induced by strong electron correlations and geometrical frustration are very sensitive to dimensionality.
In other words, geometrical frustration and low dimensionality have a key role of intriguing physics at least in our model in this study.

\section*{Acknowledgment}

This work was partly supported by a Grant-in-Aid for Scientific Research on Priority Areas from the Ministry of Education,
Culture, Sports, Science and Technology, Japan (MEXT), and also by a Next Generation Supercomputing Project, Nanoscience Program, MEXT, Japan. 

A part of the computation was done at the supercomputer center in ISSP, University of Tokyo.

\appendix

\section{Calculation of the poles of the lattice Green's function}
\label{sec:appendixa}

The poles ${\omega}_{m}$ of the Green's function $({\bf G}^{-1}_{0,{\bf t}}-{\bf {\Sigma}}_{{\bf t^{\prime}},{\bf U}})^{-1}
{\verb| |}{\equiv}{\verb| |}{\bf G}_{{\bf t},{\bf U}}$  
can be obtained in the following way
{\cite{aichhorn-hightc2}}:
Consider the Lehmann representation of ${\bf G}_{{\bf t^{\prime}},{\bf U}}$ which can be cast into the form
\begin{equation}
G_{{\alpha},{\beta},{\bf t^{\prime}},{\bf U}}({\omega})={\sum_{m}}Q_{{\alpha},m}\frac{1}{{\omega}-{\omega}^{\prime}_{m}}Q^{\dagger}_{m,{\beta}},
\end{equation}
where ${\alpha}=($site $i$, spin ${\sigma})$ in the cluster.
The ``$Q$-matrix" is defined as
\begin{equation}
Q_{{\alpha,m}}\stackrel{T=0}{=}{\delta}_{r,0}{\langle}0{\mid}c_{\alpha}{\mid}s{\rangle}+{\delta}_{s,0}{\langle}r{\mid}c_{\alpha}{\mid}0{\rangle},
\label{eq:qmat1}
\end{equation}
and ${\omega}^{\prime}_{m}$ and $Q_{{\alpha,m}}$ are obtained from exact diagonalization of the cluster.
The spectral weight (residue) of $G_{{\alpha},{\beta},{\bf t^{\prime}},{\bf U}}({\omega})$ at a pole ${\omega}={\omega}^{\prime}_{m}$
is given by $Q_{{\alpha},m}Q^{\dagger}_{m,{\beta}}$.
${\mid}0{\rangle}$ denotes the (grand-canonical) ground state of the reference system.
Introducing the diagonal matrix $g_{m,n}({\omega})={\delta}_{m,n}/({\omega}-{\omega}^{\prime}_{m})$, we have:
\begin{equation}
{\bf G}_{{\bf t^{\prime}},{\bf U}}({\omega})={\bf Q}{\bf g}({\omega}){\bf Q}^{\dagger}.
\label{eq:qmat2}
\end{equation}
Defining ${\bf V}={\bf t}-{\bf t^{\prime}}$, the VCA expression for the lattice Green's function can be written as
\begin{equation} 
{\bf G}_{{\bf t},{\bf U}}{\equiv}\frac{1}{{\bf G}^{-1}_{0,{\bf t}}-{\bf {\Sigma}}_{{\bf t^{\prime}},{\bf U}}}
=\frac{1}{{\bf G}^{-1}_{{\bf t^{\prime}},{\bf U}}-{\bf V}}.
\end{equation}
This expression can be transformed with the help of the ${\bf Q}$-matrix in Eqs. ({\ref{eq:qmat1}}) and ({\ref{eq:qmat2}}):
\begin{equation}
{\bf G}_{{\bf t},{\bf U}}=\frac{1}{({\bf Q}{\bf g}{\bf Q}^{\dagger})^{-1}-{\bf V}}={\bf Q}\frac{1}{{\bf g}^{-1}-{\bf Q}^{\dagger}{\bf V}{\bf Q}}{\bf Q}^{\dagger}.
\label{eq:g1}
\end{equation}
Since ${\bf g}^{-1}={\omega}-{\bf {\Lambda}}$ with ${\Lambda}_{m,n}={\delta}_{m,n}{\omega}^{\prime}_{m}$,
the poles of ${\bf G}_{{\bf t},{\bf U}}$ are now simply given by the eigenvalues of the (frequency independent) matrix 
${\bf M}={\bf {\Lambda}}+{\bf Q}^{\dagger}{\bf V}{\bf Q}$
and can be easily found by numerical diagonalization.
The dimension of ${\bf M}$ is given by the number of poles of ${\bf G}_{{\bf t^{\prime}},{\bf U}}$ with nonvanishing spectral weight.

\section{Calculation of the supercluster Green's function}
\label{sec:appendixb}

The Green's function of the supercluster can be calculated as follows:
Switching off the hopping processes that connect the $N_{c}=4$ clusters in the supercluster
gives a block-diagonal Hamiltonian which can be treated by numerical diagonalization. 
The switched off hopping processes are then incorporated again perturbatively.
The supercluster Green's function ${\bf G}^{\rm (s.c.)}_{{\bf t}^{\prime},{\bf U}}$ can be written as
\begin{equation}  
{\bf G}^{\rm (s.c.)}_{{\bf t}^{\prime},{\bf U}}=\frac{1}{{\bf G}^{{\rm (block)}-1}_{{\bf t^{\prime}},{\bf U}}-{\bf V}^{\rm (s.c.)}},
\label{eq:g-sc}
\end{equation}
where
\begin{eqnarray}
{\bf G}^{{\rm (block)}}_{{\bf t^{\prime},{\bf U}}}&{\equiv}&
\left[ 
\begin{array}{cc}
{\bf G}^{\rm (1)}_{{\bf t^{\prime}},{\bf U}} & {\bf 0} \\
{\bf 0} & {\bf G}^{\rm (2)}_{{\bf t^{\prime}},{\bf U}} \\
\end{array} 
\right] \nonumber \\
&=&
\left[ 
\begin{array}{cc}
{\bf Q}^{\rm (1)}{\bf g}^{\rm (1)}({\omega}){\bf Q}^{{\rm (1)}{\dagger}} & {\bf 0} \\
{\bf 0} & {\bf Q}^{\rm (2)}{\bf g}^{\rm (2)}({\omega}){\bf Q}^{{\rm (2)}{\dagger}} \\
\end{array} 
\right] \nonumber \\
&=&
\left[ 
\begin{array}{cc}
{\bf Q}^{\rm (1)} & {\bf 0} \\
{\bf 0} & {\bf Q}^{\rm (2)} \\
\end{array} 
\right]
\left[ 
\begin{array}{cc}
{\bf g}^{\rm (1)}({\omega}) & {\bf 0} \\
{\bf 0} & {\bf g}^{\rm (2)}({\omega}) \\
\end{array} 
\right] \nonumber \\
&&{\times}
\left[ 
\begin{array}{cc}
{\bf Q}^{{\rm (1)}{\dagger}} & {\bf 0} \\
{\bf 0} & {\bf Q}^{{\rm (2)}{\dagger}} \\
\end{array} 
\right] \nonumber \\
&{\equiv}&
{\bf Q}^{\rm (block)}{\bf g}^{\rm (block)}({\omega}){\bf Q}^{{\rm (block)}{\dagger}}.
\label{eq:g-block}
\end{eqnarray}
Here, superscripts (1) and (2) denote a cluster index in a supercluster.
${\bf V}^{\rm (s.c.)}$ is the intercluster hopping between the two clusters.
The expression Eq. ({\ref{eq:g-sc}}) can be transformed with the help of the ${\bf Q}^{\rm (block)}$-expression 
of ${\bf G}^{{\rm (block)}}_{{\bf t^{\prime},{\bf U}}}$ of Eq. (\ref{eq:g-block}):
\begin{eqnarray}
{\bf G}^{\rm (s.c.)}_{{\bf t}^{\prime},{\bf U}}&=&\frac{1}{({\bf Q}^{\rm (block)}{\bf g}^{\rm (block)}{\bf Q}^{{\rm (block)}{\dagger}})^{-1}-{\bf V}^{\rm (s.c.)}} \nonumber \\
&=&{\bf Q}^{\rm (block)}\frac{1}{{\bf g}^{{\rm (block)}{-1}}-{\bf Q}^{{\rm (block)}{\dagger}}{\bf V}^{\rm (s.c.)}{\bf Q}^{\rm (block)}} \nonumber \\
&&{\times}{\bf Q}^{{\rm (block)}{\dagger}}.
\end{eqnarray}
Since ${\bf g}^{{\rm (block)}{-1}}={\omega}-{\bf {\Lambda}}^{\rm (block)}$ where ${\bf {\Lambda}}^{\rm (block)}$ 
is the diagonal matrix the elements of which are the poles of ${\bf g}^{\rm (1)}({\omega})$ and ${\bf g}^{\rm (2)}({\omega})$,  
the poles of ${\bf G}^{\rm (s.c.)}_{{\bf t}^{\prime},{\bf U}}$ are given by the eigenvalues of the matrix 
${\bf M}^{\rm (s.c.)}={\bf {\Lambda}}^{\rm (block)}+{\bf Q}^{{\rm (block)}{\dagger}}{\bf V}^{\rm (s.c.)}{\bf Q}^{\rm (block)}$
and can be found by numerical diagonalization.
Diagonalizing ${\bf M}^{\rm (s.c.)}$ by an appropriate matrix ${\bf P}$ as
\begin{equation}
{\bf M}^{\rm (s.c.)}={\bf P}{\overline{\bf M}}^{\rm (s.c)}{\bf P}^{\dagger},
\end{equation}
where ${\overline{\bf M}}^{\rm (s.c)}$ is the diagonal matrix the elements of which are the poles of ${\bf G}^{\rm (s.c.)}_{{\bf t}^{\prime},{\bf U}}$,
${\bf G}^{\rm (s.c.)}_{{\bf t}^{\prime},{\bf U}}$ can be expressed as
\begin{equation}
{\bf G}^{\rm (s.c.)}_{{\bf t^{\prime}},{\bf U}}({\omega})={\bf Q}^{\rm (s.c.)}{\bf g}^{\rm (s.c.)}({\omega}){\bf Q}^{{\rm (s.c.)}{\dagger}},
\label{eq:g-sc2}
\end{equation}
with 
\begin{eqnarray}
{\bf g}^{\rm (s.c.)}({\omega})&=&\frac{1}{{\omega}-{\overline{\bf M}}^{\rm (s.c)}}, \\
{\bf Q}^{\rm (s.c.)}&=&{\bf Q}^{\rm (block)}{\bf P}, \\
{\bf Q}^{{\rm (s.c.)}{\dagger}}&=&{\bf P}^{\dagger}{\bf Q}^{{\rm (block)}{\dagger}}.
\end{eqnarray}
Since Eq. ({\ref{eq:g-sc2}}) has the same form as Eq. ({\ref{eq:qmat2}}), 
one repeat the procedure from Eq. ({\ref{eq:qmat2}}) to Eq. ({\ref{eq:g1}}) once again
in order to obtain the poles of the lattice Green's function.

\end{document}